\documentclass{hotnets2020}
\usepackage{amsmath,amssymb,amsfonts}
\usepackage{algorithmic}
\usepackage{commath}
\usepackage{graphicx}
\usepackage{mathtools}
\usepackage{textcomp}
\usepackage{xcolor}
\usepackage[sorting=none]{biblatex}

\usepackage{times}
\usepackage{hyperref}

\usepackage[disable]{todonotes}

\title{TCP D*: A Low Latency First Congestion Control Algorithm }

\author{Taran Lynn\\
\textit{Computer Science} \\
\textit{University of California, Davis}\\
Davis, California \\
https://orcid.org/0000-0003-4412-7612\\
\and
Dipak Ghosal\\
\textit{Computer Science} \\
\textit{University of California, Davis}\\
Davis, California \\
\url{ghosal@cs.ucdavis.edu}
}

\newcommand{\sndcwnd}{\text{snd\_cwnd}}
\newcommand{\deliv}{\text{Delivered}}
\newcommand{\gaincwnd}{\text{gain\_cwnd}}
\newcommand{\inflight}{\text{Inflight}}
\newcommand{\rtt}{\text{RTT}}
\newcommand{\bdp}{\text{BDP}}
\newcommand{\drate}{\text{Delivery Rate}}
\newcommand{\prate}{\text{Pacing Rate}}

\newcommand{\mode}{\text{Mode}}
\newcommand{\sincemode}{\text{Time since mode change}}
\newcommand{\last}[1]{\text{last}~#1}

\newcommand{\figurename}{Figure}
\newcommand{\tablename}{Table}

\begin{document}

\maketitle

\begin{abstract}
  The choice of feedback mechanism between delay and packet loss  has long been a point of contention in TCP congestion control. This has partly been resolved, as it has become increasingly evident that delay based methods are needed to facilitate modern interactive web applications.
  However, what has not been resolved is what control should be used, with the two candidates being the congestion window and the pacing rate. BBR is a new delay based congestion control algorithm that uses a pacing rate as its primary control and the congestion window as a secondary control.
  We propose that a congestion window first algorithm might give more desirable performance characteristics in situations where latency must be minimized even at the expense of some loss in throughput. To evaluate this hypothesis we introduce a new congestion control algorithm called TCP D*, which is a congestion window first algorithm that adopts BBR's approach of maximizing delivery rate while minimizing
    latency. In this paper, we discuss the key features of this algorithm, discuss the differences and similarity to BBR, and present some preliminary results based on a real implementation. 
\end{abstract}



\section{Introduction}

TCP congestion control has been a long studied problem in networking~\cite{jacobson1988congestion, blanton2009rfc5681, kurose2010computer}.
The first step in understanding this complex issue is to define and
quantify what congestion is. Congestion can generally be understood as the degradation in  network performance as packets are transmitted faster than the rate at which parts of the network (the bottlenecks) can process them.
We quantify the congestion by considering the outstanding queues at these bottlenecks and  we can equate the level of congestion to the number of packets in the queue. As congestion increases each packet must wait longer in the queue, increasing latency. Additionally, if the number of packets exceeds the maximum queue length, then they are dropped.

Obviously, dropped packets are bad because they need to be re-transmitted for TCP to achieve reliable delivery. Congestion also degrades the goodput of the network and increase latency, which poses other major issues.
The first major issue is user experience. Many modern network applications such as AR/VR, online gaming, video
conferencing, and remote systems administration are sensitive to
latency change on the order of a few 10s of milliseconds~\cite{nikravesh2016depth,elbamby2018toward}.
If latency degrades far enough, even web browsing is impacted. Empirical study shows that the QoE of web browsing can quickly deteriorate with delay greater than 2~seconds~\cite{zhang2019e2e}.
The second issue occurs when multiple flows  from heterogeneous sources are present.
Specifically if some sources are application limited, while others are not,
then increased latency from congestion can cause the smaller flows to be starved of throughput. For example, if there is one flow that is downloading a large file (such as an hour long movie), then without limiting throughput to prevent congestion, smaller flows (like web page requests) can take much
longer to download~\cite{gettys2011bufferbloat}.

Now we understand why we want 
to reduce congestion, and preferably we will have no
packets being queued. Additionally, it is important to ensure that the network links are fully utilized. 
\textbf{Thus we can state the goal of congestion control as such,
maximize throughput while minimizing latency.}
Ideally we would send data as fast as the bottleneck can process it, while at the same time maintaining an empty queue.
This is known as Kleinrock's optimal operating point~\cite{kleinrock2018internet}. This optimal operating point is achieved by maximizing power, which is the ratio of the throughput to the delay, thus achieving maximum throughput with the minimum delay. 

The  oldest and yet the most prevalent methods of  congestion control use losses as the primary feedback of congestion~\cite{jacobson1988congestion, blanton2009rfc5681, cubic}. This is far from ideal, as losses only occur when bottleneck queues are full and congestion is maximized.
Things get even worse with deep buffers at the bottleneck links. 
In the worst case this can lead to large latencies, on the order of
seconds in poorly configured networks.
This effect has been termed bufferbloat~\cite{gettys2011bufferbloat}.
One way to mitigate this is through active queue management (AQM)~\cite{nichols2012controlling, aqm-pie, floyd1993random}.
However, adoption of AQM on commodity devices has been slow.

A better alternative to loss based congestion control is delay based congestion
control~\cite{TCP-Vegas,bbr_tcp,timely}.
Delay based algorithms use the  packet's
\textbf{round-trip times} (RTTs), to estimate latency.
Since there is a direct relationship between congestion and RTT, delay
based algorithms can operate much closer to Kleinrock's optimal operating
point. There are also congestion control algorithms that are hybrids in that they use both delay and loss as feedbacks. TCP Illinios~\cite{TCP-Illinois} and TCP Compound~\cite{TCP-Compound} are two examples. 

The next consideration in congestion control is how to control the
rate at which packets are sent and how many packets are in queues.
The most straightforward way to do this it to set a \textbf{pacing rate},
which is the rate at which a client transmits data. A pacing rate determined by the congestion control algorithm can be maintained using  a qDisc such as FQ/pacing and HTB~\cite{htb} or a scalable  traffic shaping function  implemented on a programmable NIC such as  Carousel~\cite{carousel}.

The other approach to control the rate at which packets are sent is to use a \textbf{congestion window}, which sets the maximum number of unacknowledged packets that a client may send. This is how most of the congestion algorithms such as TCP Reno~\cite{blanton2009rfc5681},  CUBIC~\cite{cubic}, and HTCP~\cite{htcp} operate.  
As the tradeoffs between the methods are  very important, we discuss them below.

When dealing with clients that send packets in bursts, a congestion
window may let packets through all at once, overwhelming bottlenecks
and leading to temporary congestion.
By pacing the packets instead we can smooth out the sending of packets
to a rate that bottlenecks can tolerate.

One advantage of the congestion window is that it has a linear
relationship to the length of any outstanding queues, and thus to the
degree of congestion.
Briefly, the physical network links will be transmitting a number of
packets equal to the network's \textbf{bandwidth-delay product} (BDP).
If the congestion window is greater than the BDP, than the excess
packets must be stored in queues.
The fact that the number of packets in queues is the difference of the
BDP and congestion window is a core principal of many congestion
control algorithms~\cite{jacobson1988congestion, blanton2009rfc5681, cubic}.
Conversely, the pacing rate does not have a linear relationship with
how many packets are in queues.
In fact, when operating close to the bottleneck rate, small changes
can lead to large fluctuations in the number of queued packets~\cite{kleinrock1975queueing, jacobson1988congestion}.
This means pacing rate based algorithms must react quickly and
precisely to changes in the network to maintain optimal operation.

In this paper we discuss a newly developed TCP congestion control
algorithm called TCP D*.
It adopts BBR's approach of maximizing delivery rate while minimizing
latency~\cite{bbr_tcp}.
However, TCP D* uses the congestion window as its primary control.
We posit that this can result in a lower RTT, although potentially at
the expense of throughput.
With this paper, we hope to open up a discussion about the trade-offs of using a
congestion window versus pacing packets.
We found that generally the congestion window approach reduces latency potentially at the expense of some loss in throughput.
On the other hand, a pacing rate approach maximizes throughput,
potentially at the expense of increased latency.

\section{Related Work}

TCP D* is built upon earlier work in RTT based congestion control
algorithms.
One of the first, and simplest, such algorithms was TCP Vegas~\cite{TCP-Vegas}.
TCP Vegas extended TCP Reno~\cite{blanton2009rfc5681} in order to react to RTT congestion
signals.
In essence, TCP Vegas used the difference between the base RTT and current
RTT to estimate the number of packets at the bottleneck, and attempts to keep it in
the range of 2-4 packets.
This resulted in a lower RTT and better throughput than TCP Reno under
good conditions.
However, TCP Vegas  performed poorly under changing network conditions,
due to maintaining the same base RTT estimate for the entire
lifetime of the flow.
Similar algorithms that use variations of this technique include 
FAST TCP~\cite{TCP-FAST}  and  Compound TCP~\cite{TCP-Compound}. 

TIMELY~\cite{timely} is a delay based algorithm that diverges significantly from TCP Vegas.
TIMELY uses the \textbf{delay gradient}, or the derivative of the RTT, to
control the pacing rate. In particular, it increases the rate on negative gradients, and decreases the rate on
positive gradients.
Delay gradient computation requires very accurate RTT values, which
can only be realistically obtained using hardware support in the NIC.
This makes TIMELY more suitable to data center use.

TCP LoLa~\cite{TCP-LoLa} is another delay based algorithm that uses a congestion window.
TCP LoLa's goal is to set the RTT to a fixed amount over the base RTT,
which hopefully allows it to fully utilize throughput while keeping latency
at tolerable levels.
To do this it cycles between a cubic gain cycle, a fair flow balancing
cycle, and a holding cycle.

\subsection{Relation to BBR}

TCP D*  is most closely related to BBR, and requires the most thorough
comparison.
The clearest way to do this is to compare their similarities and
differences.

\subsubsection{Similarities}

\begin{itemize}
\item Both try to operate near Kleinrock's operating point that occurs
  when the ratio of the throughput to the RTT is maximized.
  This implies that throughput is set to the maximum bandwidth, while
  the RTT is set to the base RTT.

\item Both use packet delivery rate and RTT as primary feedback  signals.

\item Both use modal operation.
  They transition between modes where they operate at Kleinrock's
  optimal operating point and where they probe for more bandwidth or
  a lower base RTT.
\end{itemize}

\subsubsection{Differences}

\begin{itemize}
\item TCP D* uses the send congestion window as it's primary control,
  while BBR  uses the pacing rate as the primary control, with the congestion window
  as a secondary control.

\item The gain in TCP D*'s control algorithm is based on the variance of the BDP estimate.
  On the other hand, BBR uses a constant multiple of its bottleneck rate estimate.

\item TCP D* sets the window to its estimated BDP, while BBR sets its
  window to twice its estimated BDP.
  This means that in the worst case, wherein  BBR's pacing rate control
  is not enough to prevent congestion, BBR may have an RTT twice that of
  TCP D*.
\end{itemize}

%

\section{Implementation}

TCP D* is implemented as a Linux kernel module.
It consists of both a congestion avoidance and a slow start algorithm.
Before discussing these algorithms in detail, it is important to
understand the basic principles behind TCP D*.

\subsection{First Principles}

TCP D* uses the sending congestion window as its primary control.
We use a windowed approach because there is a direct relationship
between it and congestion.
This relationship is that if the window is larger than the
bandwidth-delay product (BDP), then the excess data must be stored in
a buffer.
The greater the amount of data stored in the buffer, the higher the
round trip time (RTT), which serves as our main feedback.
This relationship is linear and generally well behaved, which makes
the congestion window and RTT combination ideal for congestion control.

To achieve an optimal control we need to model the network.
There are four primary equations that relate the number of delivered
bytes, RTT, BDP, and the delivery rate of data.
The first defines the straight forward relationship for the delivery rate.
For any flow indexed by $i$, we have
\begin{equation}
  \drate_i(t) = \dod{}{t} \deliv_i(t) . \label{eq:delivery_rate}
\end{equation}
The second equation  defines the BDP for the flow as
\begin{equation}
  \bdp_i(t) = (\drate_i(t)) (\min\rtt_i) . \label{eq:bdp}
\end{equation}
The third equation simply states that the RTT must be at least its
minimum value, or
\begin{align}
  \rtt_i(t) &\geq \min\rtt_i . \label{eq:min_bound}
\end{align}
The final and most important equation relates the RTT to all flows
through a bottleneck, and is
\begin{align}
  \rtt_i(t) &= \min\rtt_i + \frac{\sum_{i = 1}^N (\inflight_i(t) - \bdp_i(t))}{\sum_{i = 1}^N \drate_i(t)} . \label{eq:model}
\end{align}
To justify this consider a packet being sent through the network.
The packet must at least experience the base RTT from propagation delay,
giving the $\min\rtt$ expression.
Additionally, if the number of in-flight bytes is greater than the
BDP, then the extra packets must be buffered.
This buffer can only be emptied at the delivery rate and we get the
rest of the equation.
Also note that Eqs.~ \eqref{eq:bdp}, \eqref{eq:min_bound}, and
\eqref{eq:model} imply that
\begin{align}
  \sum_{i = 1}^N (\drate_i(t)) (\min\rtt_i) &\leq \sum_{i = 1}^N \inflight_i(t) .
\end{align}
This means that if the delivery rate is not restricted by a
bottleneck, then it is determined by the number of in-flight bytes.
When the two sides are unequal the additional in-flight bytes are no
longer contributing to throughput, and is only increasing latency.
Thus, our goal is to find the maximum $\inflight_i(t)$ such that
$\inflight_i(t) = \bdp_i(t)$, at which point the delivery rate is
maximized and $\rtt_i = \min\rtt_i$ from Eq.~\eqref{eq:model}.

\subsection{Main Algorithm}

The pseudocode for the main algorithm is given in
\figurename~\ref{fig:alg_main}.
There are four important modes the algorithm cycles between.
In all of the following modes the pacing rate is unlimited, with the
congestion window acting as the only control.

\subsubsection{DRAIN} The first mode to be run after slow start is DRAIN.
In normal operation it is run whenever $\min\rtt$ has not been updated
for at least 10 seconds.
The purpose of this mode it to re-estimate the $\min\rtt$ and thus
adapt to changing network conditions.
To achieve this goal we drain the pipe of all but the minimum number
of packets necessary for feedback.

\subsubsection{GAIN 1}
The main purpose of GAIN 1 is to prep the network pipe for bandwidth
estimation.
From the results derived from Eq.~\eqref{eq:bdp}, Eq.~\eqref{eq:min_bound},
and Eq.~\eqref{eq:model}, we know that we want to set the congestion window
equal to the maximum BDP.
However, if we set it equal to the current BDP there is no way to
detect unused throughput.
We thus set the congestion window slightly above the BDP (with
$\sndcwnd = \bdp + \gaincwnd$) so that we can detect unused
throughput.
We then start tracking the number of delivered packets and enter GAIN
2.

\subsubsection{GAIN 2}
GAIN 2 computes the delivery rate given by Eq.~\eqref{eq:delivery_rate}
and uses it and the $\min\rtt$ estimation to calculate the new BDP
estimate given by Eq.~\eqref{eq:bdp}.
It also sets the $\gaincwnd$ (see below).
Note that the delivery rate is only tracked in GAIN 2, but not GAIN 1.
This is done to allow time for the feedback to adjust when we change the congestion window.
Immediately tracking delivery rate after such a change would result in under or over
estimates, depending on if the window increased or decreased.
At the end of the mode we enter DRAIN if a $\min\rtt$ timeout has
occurred, otherwise we fall back to GAIN 1.

\subsubsection{$\gaincwnd$}
The $\gaincwnd$ determines how we increase $\sndcwnd$.
Ideally, we want the $\gaincwnd$ to be as small as possible at steady
state to minimize overhead, but large enough so that under-utilized
bandwidth is quickly reclaimed.
While using a multiplicative increase (MI) could achieve these goals, we
have found that this approach results in unfairness between flows.
In order to maintain good fairness while in steady state, $\gaincwnd$
must stay the same or decrease as the $\sndcwnd$ increases.
This rules out MI.

Using twice the absolute deviation of our BDP estimates was
found to balance all of these requirements.
When reclaiming bandwidth it results in exponential growth.
This is because when we are gaining bandwidth the deviation in the BDP
is equal to the last $\gaincwnd$, which ends up doubling the
$\gaincwnd$ every round.
However, in steady state the $\gaincwnd$ corresponds to the variance in the throughput,
which is generally low relative to the overall throughput and roughly the same for all flows.
Note that in the actual code we limit the $\gaincwnd$ to the range $[4, \bdp]$.
This is done to ensure that the algorithm continues probing at low variations, and does not over-react to large variations.

\begin{figure}[h]
  \begin{algorithmic}
    \STATE{$\prate \leftarrow \infty$}

    \LOOP

    \STATE{$\min\rtt \leftarrow \min\cbr{\min\rtt, \rtt}$}
    \STATE{}

    \IF{$\mode = \text{DRAIN}$}
      \IF{$\sincemode > 2 \cdot \rtt$}
        \STATE{$\sndcwnd \leftarrow \bdp + \gaincwnd$}
        \STATE{Switch to GAIN 1}
      \ENDIF
    \ELSIF{$\mode = \text{GAIN 1}$}
      \IF{$\sincemode > 2 \cdot \rtt$}
        \STATE{Start tracking of delivered packets}
        \STATE{Switch to GAIN 2}
      \ENDIF
    \ELSIF{$\mode = \text{GAIN 2}$}
      \IF{$\sincemode > \rtt$}
        \STATE{$\last\bdp \leftarrow \bdp$}
        \STATE
        \STATE\COMMENT{From \eqref{eq:delivery_rate}}
        \STATE{$\drate ~\coloneqq~ \frac{\deliv}{\sincemode}$}
        \STATE\COMMENT{From \eqref{eq:bdp}}
        \STATE{$\bdp \leftarrow (\drate) (\min\rtt)$}
        \STATE
        \STATE{$\sigma ~\coloneqq~ \abs{\bdp - \last\bdp}$}
        \STATE{$\gaincwnd \leftarrow 4 + \min\cbr{2\sigma, \bdp}$}

        \STATE
        \IF{Time since $\min\rtt$ was last updated $>$ 10 seconds}
          \STATE{$\sndcwnd \leftarrow 4$}
          \STATE{$\min\rtt \leftarrow \rtt$}
          \STATE{Switch to DRAIN}
        \ELSE
          \STATE{$\sndcwnd \leftarrow \bdp + \gaincwnd$}
          \STATE{Switch to GAIN 1}
        \ENDIF
      \ENDIF
    \ENDIF

    \ENDLOOP
  \end{algorithmic}

  \caption{The pseudo-code of TCP D*.}
  \label{fig:alg_main}
\end{figure}

\subsection{Slow Start}

Another important part of the algorithm is its slow start routine.
Slow start is entered on two occasions, those being when transmission
starts and when the flow is restarted after a timeout \cite{rfc2861}.
The slow start routine is essentially the same as the main algorithm,
with two major exceptions.
\begin{enumerate}
\item It never enters DRAIN mode.

\item It sets
  \begin{align}
    \gaincwnd(t) &\leftarrow \max\cbr{4, \frac{1}{2} \bdp(t)} , \label{eq:ss_gain_cwnd}
  \end{align}
  instead of what is given in \figurename~\ref{fig:alg_main}.
\end{enumerate}
These changes allow for rapid estimation of the BDP while keeping the
core attributes intact.
Slow start ends when either a loss or ECN congestion event (CE)
occurs, or when the BDP estimate stops increasing.
Note that a loss occurring is a worst case scenario.
If there is at least one BDP of upstream buffering then
Eq.~\eqref{eq:ss_gain_cwnd} should guarantee that we do not send more data
than can be buffered.

\section{Results}

The TCP D* Linux module has been tested under a wide variety of
network conditions, both physical and simulated.
These conditions include dedicated wide area networks (WANs), poorly
tuned networks that are easily congested, and networks with different
RTT flows.
Except for cases where simulation was used, we also tested BBR \cite{bbr_tcp},
CUBIC \cite{cubic}, TCP Reno \cite{jacobson1988congestion}, and TCP Vegas \cite{TCP-Vegas}.
The tests where conducted using Flent~\cite{hoiland2017flent, flent}, which is a wrapper
around the Netperf and ping testing utilities.

\subsection{Chameleon Testbed}
For our first evaluation we used Chameleon which is a configurable
experimental environment for large-scale cloud
research~\cite{keahey2020lessons}.
The Chameleon setup consisted of two sites, University of Chicago and
the Texas Advanced Computing Center (TACC), connected by a dedicated
WAN.
The defining characteristics of this case were low variance in
throughput and RTT.
It was also difficult to increase RTT by saturating the network with
packets, possibly due to intermediate shaping or the availability of
many routes for packets.
A dedicated baremetal server running Linux at each location was used
for this test.

\begin{figure}[!ht]
  \centering
  \includegraphics[width=\linewidth]{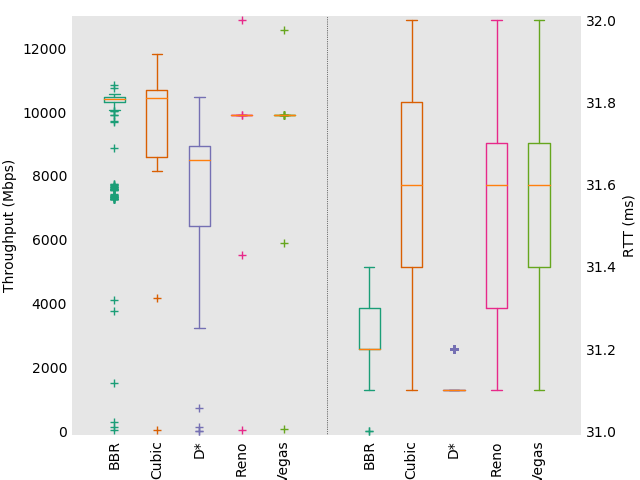}
  \caption{The achieved throughput and RTT for 1 flow between the two different sites of the Chameleon Cloud.
    TCP D* achieves the lowest RTT at the expense having lower throughput.
    A likely cause is a combination of TCP D*'s conservative congestion window growth 
    and a low variance in its BDP estimates.
    The long tail of throughput for BBR and TCP D* is due to their probing cycles.
  }
  \label{fig:chameleon_1up}
\end{figure}

\figurename~\ref{fig:chameleon_1up} compares the performance of the
congestion algorithms with a single upload stream.
TCP D* performs well in regards to RTT, but lags in throughput.
This is unique to this test case.
We believe the cause is a combination of TCP D*'s conservative window and
that window increase is controlled by variance in our BDP estimates.
For a network with low throughput variance, such as a dedicated WAN, resulting in
the algorithm being unable to utilize the full bandwidth.

\begin{figure}[!ht]
  \centering
  \includegraphics[width=\linewidth]{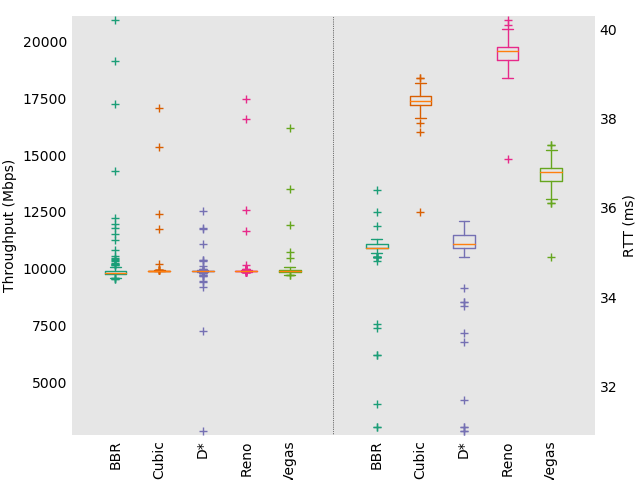}
  \caption{The achieved throughput and RTT for 128 flows between the two different sites of the Chameleon Cloud.
    BBR and TCP D* give have lowest latency.
    The tails for BBR and TCP D* are not present, because when some flows
    probe others take up the remaining throughput.}
  \label{fig:chameleon_128up}
\end{figure}

To compare how the algorithms operate under load we also tested
running 128 streams in parallel.
\figurename~\ref{fig:chameleon_128up} shows the results of these tests.
All of the algorithms used the full bandwidth, and BBR and TCP D* performed
significantly better in minimizing RTT than the other algorithms.
Additionally, \tablename~\ref{tab:chameleon_jain_index} compares how
fair the competing flows were.
It is evident that BBR and TCP D* performed much better in terms of stream
fairness than the other algorithms, with near perfect indexes.
This is a major advantage of using packet delivery rates as a
feedback, as it allows quick convergence to fair usage.
\begin{table}[t]
  \centering
  \begin{tabular}{|l|l|l|l|l|l|}
    \hline
    Algorithm & BBR & CUBIC & D* & Reno & Vegas\\ \hline  
    Jain Index & 0.9985  & 0.6237 & 0.9995 & 0.5557 &  0.7161\\
    \hline
    \hline
  \end{tabular}
  \caption{Jain fairness index~\cite{jain1999throughput} for all the algorithms for 128 flow experiment.
  BBR and TCP D* show very high degrees of fairness among flows.}
  \label{tab:chameleon_jain_index}
\end{table}

\subsection{Network Emulation using NetEm}

For the second test case we used NetEm ~\cite{hemminger2005network} to emulate a 500~Mbps network
with a 30~ms delay.
The network consisted of two commodity hardware machines (a desktop
and laptop) connected via a switch.
The bottleneck link was setup on the laptop side with a buffer size 
equal to twice the BDP, half of which was used by NetEm to emulate network
delays, and the rest used to emulate the bottleneck queue.
The desktop was then used to run Flent.
This network setup resulted in far more unpredictability and
congestion than the WAN case, and is a good proxy for a poorly
configured network.

\begin{figure}[!ht]
  \centering
  \includegraphics[width=\linewidth]{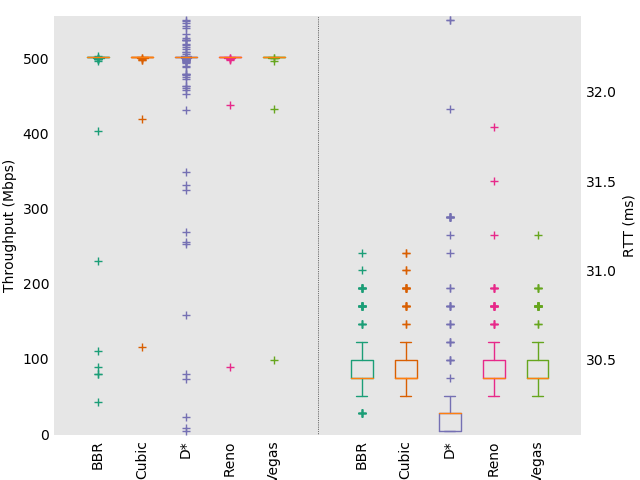}
  \caption{The achieved throughput and RTT for 1 flow on NetEm network.
    TCP D* has a marginally lower RTT than the other algorithms.
    The probing tails for BBR and TCP D* are clearly visible.}
  \label{fig:netem_1up}
\end{figure}

\begin{figure}[!ht]
  \centering
  \includegraphics[width=\linewidth]{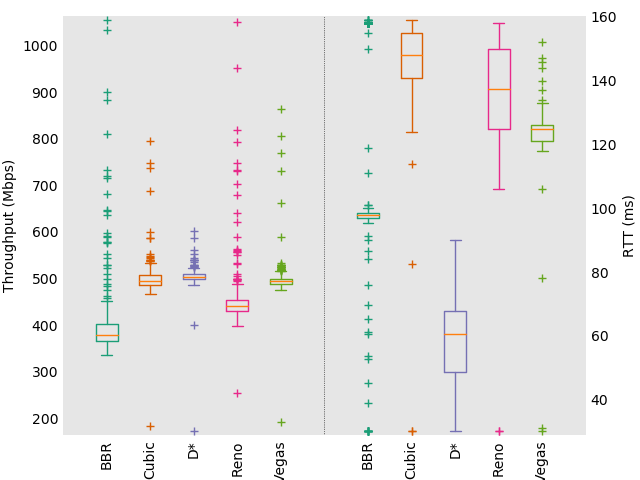}
  \caption{The achieved throughput and RTT for 128 flows on NetEm network.
  This is the test case where TCP D* shows a clear advantage, as it fully
utilizes throughput while having an RTT about 40~ms less than the next
lowest algorithm.}
  \label{fig:netem_128up}
\end{figure}

\figurename~\ref{fig:netem_1up} shows the results for one upload stream.
There is very little variation between the algorithms tested in this case.
\figurename~\ref{fig:netem_128up} shows the results for 128 upload streams.
The large number of streams resulted in serious congestion, and clear differences in performance 
between the algorithms.
In this case TCP D* gives the lowest RTT, approximately half the next
best (achieved by BBR), while using the full bandwidth.
This is in part due to TCP D* keeping the margin for congestion very
small, and its explicit modeling of congestion via the congestion
window.
It highlights TCP D*'s best use case, which is in networks with many competing flows,
especially when the intermediate links are poorly configured.

\subsection{Simulation}

\begin{figure}[!ht]
  \centering
  \includegraphics[width=\linewidth]{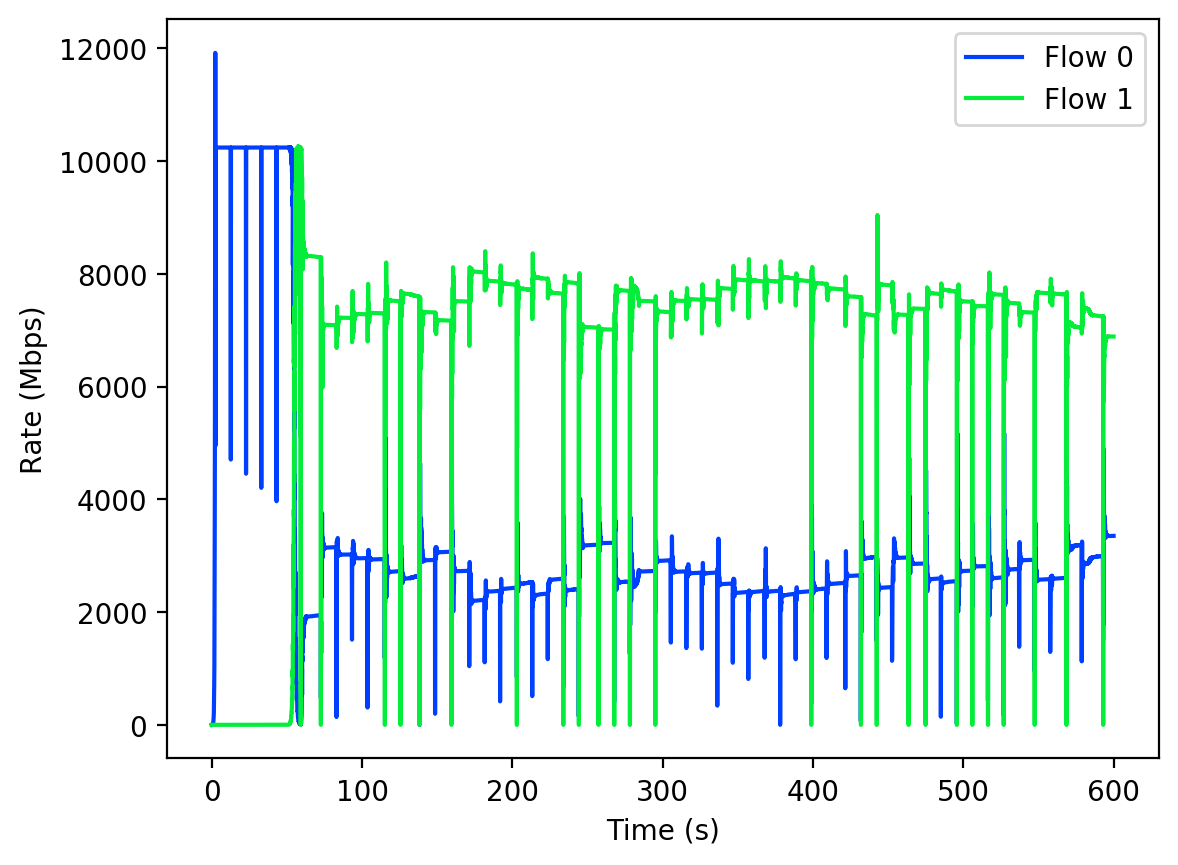}
  \caption{Rates of two simulated flows with different base RTTs.
    Flow 1 starts 10s after flow 0.
    The flows do not equally share throughput.
  Instead, each flow gets a share of throughput proportional to its RTT.}
  \label{fig:diff_rtt_rate}
\end{figure}

\begin{figure}[!ht]
  \centering
  \includegraphics[width=\linewidth]{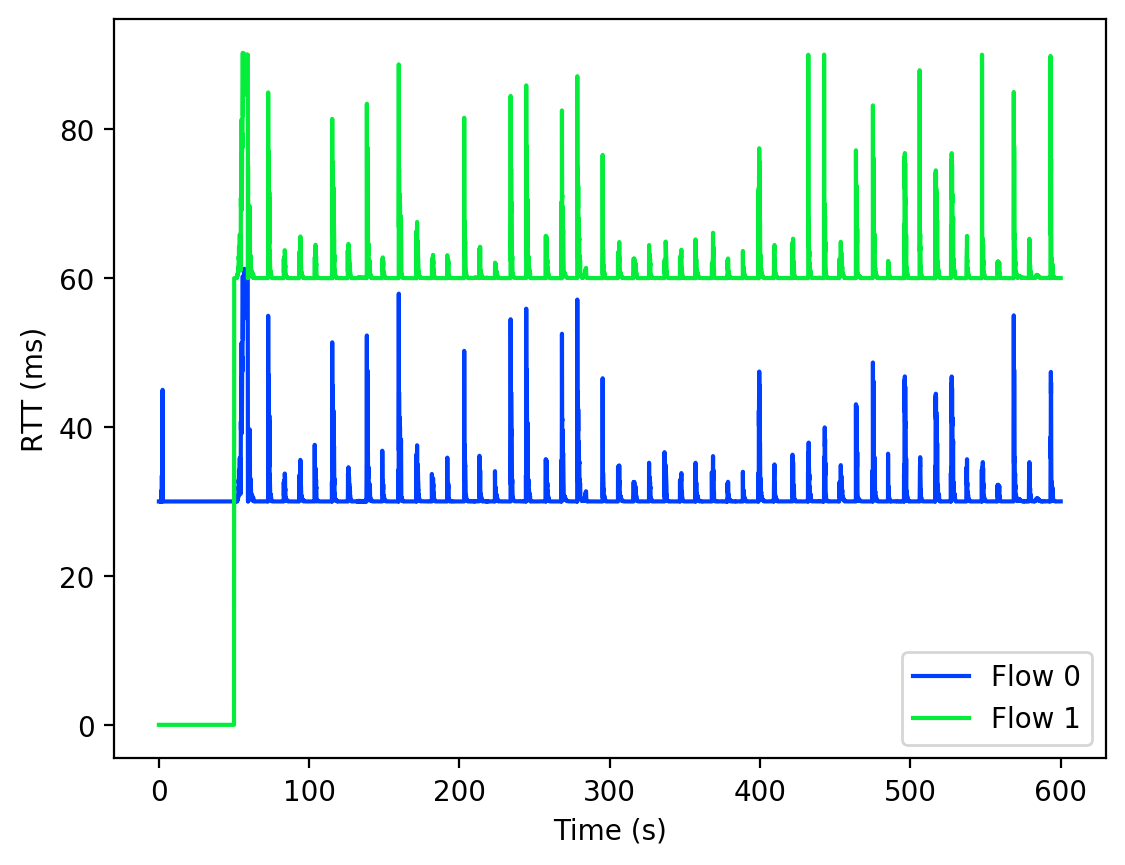}
  \caption{RTTs of two simulated flows with different base RTTs.
    Flow 1 starts 10s after flow 0.
    The base RTTs are 30~ms and 60~ms.
  Even with different RTTs both flows minimize their RTTs.}
  \label{fig:diff_rtt_rtt}
\end{figure}

The last test case was an event based simulation of two streams with
different RTTs and a common bottleneck.
This was the only test where we could not test on a physical network,
and it was thus performed using an event based simulation.
\figurename~\ref{fig:diff_rtt_rate} shows the throughput for each
stream, and \figurename~\ref{fig:diff_rtt_rtt} shows their RTTs.
The difference in throughput was roughly proportional to the
difference in each flow's base RTT.
This is in part due to that fact that the $\sndcwnd$ is set close to
the estimated BDP, which in turn is the product of the delivery rate
and base RTT.
Thus, even with a similar delivery rate, the difference in base RTTs
will result in the larger flow taking a greater share of the
bandwidth.
However, this partial sharing is still desirable, as an alternative
is one flow being completely starved.
Partial sharing thus guarantees that small flows get enough throughput
to complete in an acceptable time frame.

\section{Conclusion}

In this paper we have presented a new congestion control algorithm,
called TCP D*, which is based off of concepts from BBR, but is driven
by a congestion window first design.
Our hope is to drive discussion about the advantages and disadvantages
of using a congestion window versus a pacing rate for congestion
control.
Several questions remain to be investigated about the tradeoffs
between these approaches.
How do the two approaches perform in networks that have flows sending/receiving
data at different rates?
This is important as movie streaming and web page requests often occur
on the same network and have very different speed requirements.
How do these approaches affect performance in terms of shallow versus deep buffers?
This would be relevant for data center networks.

There are also improvements for TCP D* and inquiries into its
operation we wish to pursue in the future.
The first is to improve D*'s performance with low variance dedicated
WANs.
We have already begun working on this, and have some promising
solutions.
We will also look into how TCP D* performs under different network conditions,
specifically low-latency (sub-millisecond RTT) data center networks
and networks with packet policing or shaping.
Another improvement to consider  is  fairness when flows have different base RTTs.
A possible course of investigation is whether this can be done with a knowledge of how
many flows converge on a bottleneck, perhaps enabled by in-network telemetry.

Our goal with this paper is to  open up new avenues of research on the problem of congestion
control in new networking domains and for next generation distributed applications.
We hope that others working on congestion control will consider the lessons learned
and apply them to their own algorithms.


\pagebreak

\printbibliography

\end{document}